\documentclass[12pt]{article}
\usepackage[export]{adjustbox}
 \usepackage{mwe}
\usepackage{amsthm,amsmath,amsfonts,amssymb,fullpage}
\usepackage[authoryear]{natbib}
\usepackage[colorlinks,citecolor=blue,urlcolor=blue]{hyperref}
\usepackage[utf8]{inputenc}
\usepackage{graphicx}
\usepackage{wrapfig}
\theoremstyle{plain}

\newcommand\independent{\protect\mathpalette{\protect\independenT}{\perp}}
\def\independenT#1#2{\mathrel{\rlap{$#1#2$}\mkern2mu{#1#2}}}
\theoremstyle{remark}

\usepackage{color}
\definecolor{maroon(html/css)}{rgb}{0.5, 0.0, 0.0}

\definecolor{blue2}{rgb}{0.1, 0.2, 0.6}

\begin{document}
	
\thispagestyle{empty}
\baselineskip=28pt
\thispagestyle{empty}
\baselineskip=28pt
{\LARGE{\bf \begin{center}
A unified quantile framework for nonlinear heterogeneous 
			transcriptome-wide associations\end{center}  }}
		
		\baselineskip=12pt
		
		\vskip 2mm
		\vskip 2mm
		\begin{center}
			Tianying Wang,
			\hskip 5mm\\
			Department of Statistics, Colorado State 
			University\\
			\hskip 5mm\\
			Iuliana Ionita-Laza,
			\hskip 5mm\\
			
			Department of Biostatistics, Columbia University\\
			Department of Statistics, Lund
			University, Sweden\\
			\hskip 5mm\\
			Ying Wei				\hskip 5mm\\
			Department of Biostatistics, Columbia University
			\end{center}

	\begin{center}
		{\Large{\bf Abstract}}
		
	\end{center}
	\baselineskip=12pt

			Transcriptome-wide association studies (TWAS) are powerful tools 
			for identifying gene-level associations by integrating genome-wide 
			association studies and gene expression data.  However, most TWAS 
			methods focus on linear associations 
			between genes and traits, ignoring the complex nonlinear 
			relationships that may be present in biological systems. To address this 
			limitation, we propose a novel framework, 
			QTWAS, which integrates a quantile-based gene expression model into 
			the TWAS model, allowing for the discovery of nonlinear and 
			heterogeneous 
			gene-trait associations. Via comprehensive simulations and 
			applications to both continuous and binary traits, we 
			demonstrate that the proposed model is more powerful than 
			conventional TWAS in identifying gene-trait associations.

				\baselineskip=12pt
		\par\vfill\noindent
\par\medskip\noindent

	\clearpage\pagebreak\newpage
	\pagenumbering{arabic}
	\newlength{\gnat}
	\setlength{\gnat}{26pt}
	\baselineskip=\gnat
	\section{Introduction}\label{sec:intro}

	Over the past twenty years, genome-wide association studies (GWAS) have 
	collectively identified tens of thousands of  genetic variants associated with 
	various complex traits and 
	diseases. However, most of these variants are 
	located in non-coding regions of the genome, making it difficult to interpret 
	their functional roles.  Since it is assumed that most functional genetic 
	variants exert 
	their effects on traits through their influence on gene expression, directly 
	linking gene expression levels to phenotypes can provide a better 
	understanding of the underlying biological mechanisms and identify potential 
	therapeutic 
	targets more effectively \citep{tang2021novel,li2021tissue}. The main 
	challenge to such transcriptomic studies is that gene expression levels are 
	not easily 
	available in large-scale disease studies. To overcome this, several 
	approaches have been developed to 
	impute or predict gene expression levels based on DNA sequence data. One 
	of the most widely 
	adopted approaches is the transcriptome-wide association studies (TWAS) 
	\citep{gusev2016integrative,gamazon2015gene,zhao2021transcriptome,wainberg2019opportunities}.
	TWAS leverages  large-scale data on both genotype and gene expression 
	across various human tissues as available in projects such 
	as the Genotype-Tissue Expression (GTEx) project \citep{gtex2020gtex} to 
	learn the relationship 
	between gene expression and genetic variation 
	\citep{gusev2016integrative,gamazon2015gene,zhao2021transcriptome,wainberg2019opportunities}.
	By integrating this model with GWAS data, TWAS links predicted gene 
	expression levels to  traits of interest, providing insights into gene-trait 
	associations.

\begin{figure}
		\centering
		\includegraphics[scale = 0.45]{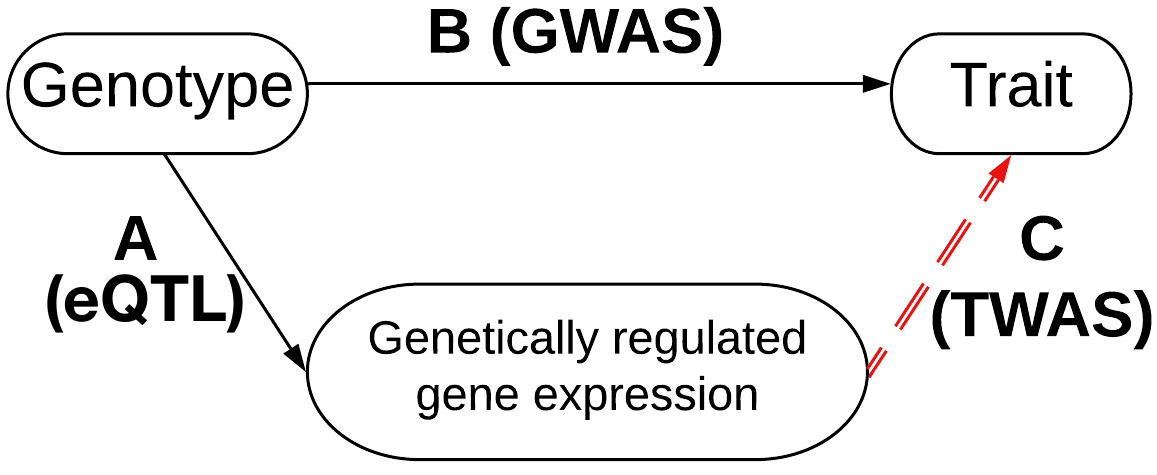}
		\caption{Quantile TWAS and classical linear TWAS models. Model A: a 
			model for SNP-gene expression association (eQTL model based on 
			GTEx data). Model B:  a model for SNP-Trait associations (GWAS 
			model). Model C: a model for expression-trait association (TWAS 
			model).}\label{fig:illustration}
	\end{figure}

	As illustrated in Figure \ref{fig:illustration}, TWAS combines two 
	distinct models --  a gene expression model (Model A) that models  gene 
	expression as a function of expression quantitative trait loci (eQTLs, i.e., 
	genetic variants that are 
	associated with gene expression), and a GWAS model (Model B ) that 
	captures the associations between a trait and individual genetic variants. The 
	two models are estimated separately and 
	then combined to infer associations between genetically regulated gene 
	expression levels and  phenotypes (Model C). For example, the widely-used 
	PrediXcan \citep{gamazon2015gene} first uses a sparse linear model such as 
	elastic net or lasso  to estimate the cis-eQTL effects on gene expression 
	and then imputes gene expression levels. Then, in the second step, it 
	formally tests the  association between imputed gene expression 
	(genetically regulated gene expression levels) and the trait of interest. 
	However access to individual-level GWAS data is often restricted due to 
	privacy concerns and data-sharing limitations. Instead, GWAS summary 
	statistics (estimated effect sizes and their standard errors) are more readily 
	available. These summary statistics not only facilitate easy access to GWAS 
	results but also 
	allow researchers to integrate data from multiple GWAS, resulting in more 
	powerful analyses. Along this direction, S-PrediXcan 
	\citep{barbeira2018exploring} extends PrediXcan to situations where only 
	GWAS summary statistics are available. More recently, 
	\cite{gusev2016integrative} 
	and \cite{nagpal2019tigar} employed Bayesian gene expression models, while 
	UTMOST \citep{hu2019statistical} introduced a multi-task learning approach 
	to jointly model gene expression across tissues. Given the practicality of 
	working with summary statistics, our proposed framework also focuses on 
	scenarios where only GWAS summary statistics are accessible.
	
	Current TWAS methodologies rely on the assumption that both the gene 
	expression model (Model A, SNP-gene expression association) and the 
	GWAS model (Model B, SNP-trait association) follow linear relationships. 
	However, multiple lines of evidence suggest substantial heterogeneity in 
	gene expression patterns, driven by genetic variation, cellular and molecular 
	diversity, and environmental, demographic, and technical factors 
	\citep{leek2007capturing,somel2006gene,budinska2013gene}. Moreover, 
	how eQTLs regulate gene expression can be highly context-dependent, 
	influenced by factors such as gene-gene interactions and gene-environment 
	interactions (GEI) \citep{umans2021disease}, leading to heterogeneous eQTL 
	effects. Therefore, linear and mean-based gene-expression models can be
	inadequate for capturing the complexity of SNP-gene expression relations. 
	Recent work by \cite{lin2022accounting}  considered a quadratic gene 
	expression model in TWAS, and observed improved power in identifying 
	gene-trait association.

	Our strategy is based on quantile regression \citep{KoenkerBassett1978} 
	which  models conditional quantiles across multiple quantile levels, and is  an 
	effective 
	approach to capture heterogeneous genetic associations 
	\citep{song2017qrank,wang2022integrated}. The flexibility of quantile 
	regression allows it to more effectively capture the variability in gene 
	expression patterns driven by genetic and environmental factors 
	\citep{wang2024genome}. In our proposed Quantile TWAS (QTWAS) 
	framework,  we use quantile regression along with a quantile-specific variable 
	screening scheme to model the entire conditional distribution of gene 
	expression given 
	the underlying genotype profile, which helps capture the heterogeneity in 
	eQTL-regulated gene expressions. 
	
	
	As we demonstrate in Appendix Section 3.1, when SNP-gene expression 
	associations are non-linear, heteroskedastic, or exhibit heavy tails, the 
	resulting gene-trait associations (Model C) can become non-linear even if 
	the GWAS associations themselves are linear. This type of non-linearity has 
	not been adequately explored in existing TWAS literature.   Allowing 
	potentially nonlinear gene-trait associations could enhance the detection 
	power, provide more nuanced insights into gene-trait associations and help 
	refine targeted interventions. Our proposed QTWAS framework addresses 
	this gap, by designing a new integration strategy that combines the 
	conditional quantiles of genetically 
	regulated gene expression with GWAS summary statistics to effectively 
	capture non-linear TWAS associations (Model C in Figure \ref{fig:illustration}). 
	By doing so, we observe significantly improved power in detecting gene-trait 
	associations. 	
	
	
	Our approach differs from existing TWAS methods in two fundamental ways. 
	First, we employ quantile regression with quantile-specific screening to model 
	the full conditional distribution of gene expression. Second, our unique 
	integration of conditional gene expression quantiles with GWAS summary 
	statistics accommodates non-linear TWAS associations.
	We also establish a theoretical framework and develop rigorous statistical 
	inference tools. Our numerical studies indicate that
	by relaxing the assumption of linear gene-trait associations, QTWAS 
	outperforms the traditional linear model-based TWAS in terms of statistical 
	power.  We applied the proposed QTWAS framework to summary statistics 
	from several GWAS 
	datasets, including continuous traits (low-density lipoprotein) from the UK 
	Biobank and binary traits (schizophrenia) where we confirm the increased 
	power of QTWAS over traditional TWAS approaches. In particular, we show 
	that the unique genes identified by QTWAS have functional enrichments 
	among 
	gene sets relevant for the corresponding traits/diseases, and furthermore 
	these 
	genes are more likely to exhibit heterogeneity in the eQTL model (Model A in 
	Figure \ref{fig:illustration}), further highlighting the advantages of the QTWAS 
	approach.

	\section{Methodology}\label{sec:method}
	
	\subsection{Notations and background}\label{sec:notations}
	We denote by $X$ the gene expression level of a 
	target gene, by $Y$ the trait/phenotype, and by $Z= (Z_1,\cdots, Z_p)^\top$ 
	a vector of $p$ genetic variants. For now we assume that $Y$ is 
	continuous but will relax this assumption later.
	SNPs $Z_1,\cdots, Z_p$ are discrete random variables with possible 
	values of ${0,1,2}$, signifying the number of reference alleles at a locus. 
	As with other TWAS methods we focus on cis-QTLs, i.e. SNPs located within 
	$\pm$1Mb from the target gene, as such regions encompass most of the 
	identified eQTLs for a gene \citep{gtex2020gtex}. Possible confounders 
	in genetic association studies including race and ethnicity, principal 
	components (PCs) of genotype data, and probabilistic estimation of 
	expression residuals (PEER) factors are normally used as covariates in 
	linear regression models to remove their effects 
	\citep{stegle2012using,hu2019statistical, stegle2010bayesian}. Without 
	loss of generality, we assume that these confounding effects on 
	phenotypes $Y$ and gene expression $X$ have been removed.  
	Furthermore, we use $C$ to represent other covariates that are uncorrelated 
	to $Z$, 
	such as age, gender, and other postnatal non-genetic factors. 
	
	\medskip
	
	As we introduced in Section \ref{sec:intro}, TWAS aims to identify the gene 
	($X$) - trait ($Y$) associations by integrating two separate models -- a 
	GWAS model for SNP-trait association, and a gene expression model 
	capturing 
	SNP-gene expression relationships. Since we rely solely on GWAS summary 
	statistics, we inherit the GWAS linear model from which the summary 
	statistics are derived.  Specifically, a typical GWAS study assumes the 
	following model: \begin{eqnarray}\label{eq:gwas}
		{\rm Genotype-Trait\ (Model\ B):}\   \   Y = \alpha_0+Z_j\beta_{\rm GWAS, 
			j}+C^\top\eta+e,
	\end{eqnarray}
	where $\alpha_0$ is the intercept, $\beta_{\rm GWAS, j}$ and $\eta$ are 
	the coefficients regarding the $j$th SNP and covariates, and $e$ is the 
	random error. We use the GWAS summary statistics, including the estimated 
	$\beta_{\rm GWAS, j}$ and their standard errors derived from this model.
	
	\medskip
	
	To fully capture the heterogeneity in genotype - gene expression 
	relationships, we model the conditional quantile of gene 
	expression $X$, denoted as  $Q_{X}(\tau\mid Z, C)$, as:
	\begin{eqnarray}\label{eq:rq_model_method}
		\lefteqn{\rm Genotype-Gene Expression\ (Model\ A):} \nonumber\\   
		Q_{X}(\tau\mid Z, C) &=& \alpha_{0}(\tau)+C^\top\alpha(\tau)+ Z^\top 
		\beta(\tau) \ \ {\rm for\ all} \ \tau \in (0,1),
	\end{eqnarray}
	where $\alpha_{0}(\tau)$, $\alpha(\tau)$, and $\beta(\tau) $ are 
	quantile-specific intercepts and slopes for covariates and genotypes, 
	respectively. Using the GTEx data for a specific tissue,  $\{X_i, \mathbf{Z}_i, 
	\mathbf{C}_i\}_{i=1}^{n}$ with sample size $n$, we can estimate Model 
	\eqref{eq:rq_model_method} by solving
	\begin{equation*}
		(\hat \alpha_{0,\tau}, \hat \alpha_\tau, \hat\beta_\tau) = 
		\arg\min_{\alpha_{0,\tau}, \alpha_\tau, \beta_\tau} 
		\sum_i^n\rho_\tau(X_i-\alpha_{0,\tau}-\mathbf{C}_i\alpha_\tau - 
		\mathbf{Z}_i\beta_\tau),
	\end{equation*}
	where $\rho_\tau(u) = |u|\{(1-\tau) I(u<0)+\tau I(u>0)\}$ is the quantile 
	regression check function with $u\in \mathbb R$, and $I(\cdot)$ is an 
	indicator function. In the later section \ref{subsec:vs}, we introduce a 
	quantile-specific screening procedure to further improve the prediction 
	accuracy. 
	
	\subsection{Generalized gene-trait association model}  \label{sec:model}
	Observed gene expression is influenced by both genetic and environmental 
	factors. In TWAS, the focus is on the association between a trait and 
	genetically regulated gene expression, which is denoted as $X_Z$ in this 
	paper. By 
	isolating the genetically regulated component, we reduce the noise 
	introduced 
	by environmental and other non-genetic influences. Consequently, we
	increase the power to detect gene-trait associations that are directly 
	mediated by 
	genetic variation.  In a linear gene expression model,  $X_Z$ can be explicitly 
	written as $E(X_Z|Z)=\sum_{j=1}^pZ_j\beta_j$. 
	With the assumption that $Z$ and $C$ are independent, which will be 
	discussed in detail in Section \ref{subsec:assu}, 
	our proposed quantile gene expression model  below (\ref{eq:qtwas2}) 
	implicitly assumes that the conditional distribution of $X_Z$ given $Z$ is 
	$Z^\top 
	\beta(\tau)$, and $\beta(\tau)$ is unspecified. As mentioned in the 
	Introduction, when the relationship between gene expression and eQTLs
	is heterogeneous, the gene-trait associations could be complex and 
	non-linear. To account for this complexity, we propose a generalized and 
	additive gene-trait association model: 
	\begin{eqnarray}
		{\rm New\  Gene-Trait \ (Model\ C):}\ \    Y = g_1(X_Z) + 
		g_2(C)+\epsilon_y,\label{eq:qtwas2} 
	\end{eqnarray} 
	where $g_1(\cdot)$ and $g_2(\cdot)$ are unknown functions. The primary 
	function of interest is $g_1(\cdot)$ which allows for a nonlinear association 
	between $Y$ and $X_Z$. Note that $g_1(\cdot) = 
	0$ corresponds to no gene-trait association between $X$ and $Y$.  When 
	both 
	$g_1$ and $g_2$ are linear functions eq, \eqref{eq:qtwas2}  degenerates to 
	the linear model in traditional TWAS, e.g. S-PrediXcan; see Appendix Section 
	1 for a review of traditional linear model based TWAS approaches.

	\subsubsection{Model Assumptions}\label{subsec:assu}
	There are several key assumptions underlying conventional TWAS. These 
	assumptions are 
	also assumed in the proposed QTWAS. The first assumption is that \textit{the 
		SNP set $(Z_1,\cdots, Z_p)$ is independent of the covariates $C$.} This 
	assumption is naturally guaranteed as most covariates are postnatal and 
	related to environmental (non-genetic) factors.  It enables the separation of 
	the genotype-contributed gene expression, denoted as $X_Z$, from the 
	other factors, making it possible to focus on testing the association between 
	the 
	trait $Y$ and $X_Z$ in the gene-trait models.  
	
	
	The second assumption is 
	that \textit{the SNP set $(Z_1,\cdots, Z_p)$ only affects $Y$ through $X$. } 
	This assumption implies the conditional independence between $Y$ and 
	$X_Z$ given $Z$, which leads to valid TWAS inference 
	\citep{barbeira2018exploring,hu2019statistical}. However, this 
	assumption 
		can be invalidated in practice by horizontal pleiotropy 
		\citep{van2024mr,barfield2018transcriptome}, the scenario where 
		a genetic variant may have an independent effect on multiple traits. 
		Horizontal pleiotropy can lead 
		to false positive discoveries in TWAS associations if the genetic variant has 
		an 
		independent effect on both the gene expression and the trait through 
		different mechanisms. Recently, TWAS approaches that remove this 
		assumption have been 
		proposed \citep{dong2020general,zhao2024adjusting,deng2021model}. In 
		Section 5, we outline strategies to remove this assumption for 
		QTWAS as well.
	
	
	The last assumption is on model transferability, in the sense 
	that 
		the eQTL effects on gene expression are the same in both reference 
		(GTEx) population and GWAS 
		population. This assumption may not hold perfectly in reality due to 
		differences in populations and environmental effects.

	\subsubsection{Inference on Gene-Trait association} 
	For each gene-trait pair, we aim to test a global hypothesis $H_0:g_1(\cdot) 
	=0$, indicating no gene-trait association. Although $g_1(\cdot)$ is unknown 
	and cannot be fully estimated without having individual-level GWAS data, we 
	can approximate it using a piece-wise linear function over quantile regions of 
	$X_Z$. We can show that the local slopes can be estimated and inferred by 
	combining the conditional quantile function of gene expression and GWAS 
	summary statistics.  Specifically, let $A_{k} = \{Q_{X_Z}(\tau_{k}), 
	Q_{X_Z}(\tau_{k+1})\}$ represent the interval between the $\tau_k$th and 
	$\tau_{k+1}$th quantiles of $X_Z$, with $\cup_k A_k$ covering the full range 
	of $X_Z$. The approximation of $g_1(\cdot)$ is given by
	\begin{eqnarray}\label{eq:linear_approx}
		g_1(X_Z)\approx\sum_{k=1}^K \gamma_{k} X_Z I\{X_Z\in A_{k}\},
	\end{eqnarray}
	where $I(\cdot)$ is an indicator function. Then testing the gene-trait 
	association is 
	equivalent to testing the null hypothesis: 
	\[
	H_0: \gamma_k = 0 \ {\rm for} \ k = 1,...,K; \ \ H_a: {\rm at\ least \ one}\  
	\gamma _k \neq 0.
	\]
	One can view $ \gamma_k$ as a population minimizer such that  $ \gamma_k 
	= \arg\min_{\boldsymbol{\gamma}} \mathbb{E}_Y(Y-\gamma_k 
	X_Z)^2I\{X_Z\in A_{k}\}$. Thus, the slope coefficient $\gamma_k$ 
	summarizes the local gene-trait association within a quantile sub-region 
	of $X_Z$, which can be written as 
	\begin{equation}
		\gamma_k=\frac{cov(X_Z, Y\mid X_Z \in A_k)}{var(X_Z\mid X_Z \in A_k)}, 
		\text{ for $k = 1,...,K$}.  \label{eq:gammakdef_method}
	\end{equation} 
	The above eq. \eqref{eq:gammakdef_method} implies that $A_k$ cannot be 
	too small; otherwise, the estimation of $var(X_Z\mid X_Z\in A_k)$ will be 
	highly unstable. Meanwhile, $A_k$ cannot be too large as that would lead 
	to a poor estimation of the unknown function $g_1(\cdot)$ according to eq. 
	\eqref{eq:linear_approx}. Our recommendation for choosing $A_k$ 
	empirically is discussed in Section \ref{sec:implement}.
	
	
	\paragraph{Estimating $\gamma_k$ through model integration} 
	$\gamma_k$ can be estimated by leveraging a conditional quantile process 
	model of the gene expression and 
	GWAS summary statistics.  We first decompose the covariance  
	$cov(X_Z,Y\mid X_Z\in A_k)$ in eq. \eqref{eq:gammakdef_method} by the 
	law of total variance:
	\begin{eqnarray}\label{eq:covEG_method}
		cov(X_Z,Y\mid X_Z\in A_k)&=&\mathbb{E}\{cov(X_Z, Y\mid Z, X_Z\in 
		A_k)\}\nonumber\\
		&&+cov\{\mathbb{E}(X_Z\mid Z, X_Z\in A_k), \mathbb{E}(Y\mid Z, X_Z\in 
		A_k)\}.
	\end{eqnarray}
	The second model assumption outlined in Section \ref{subsec:assu}  implies 
	the conditional independence 
	$Y\independent{X_Z} \mid Z$, which further implies
	$\mathbb{E}\{cov(X_Z, Y\mid Z, X_Z\in A_k)\} = 0$, and 
	$\mathbb{E}(Y\mid Z, X_Z\in A_k) = Z^\top\beta_{\rm GWAS}$, where 
	$\beta_{\rm GWAS}$ is the SNP-level effect size.
	
	
	For any continuous random variable $X$, its quantile function denoted as 
	$Q_X$ has the property  that $Q_X(U) \overset{dist.}{=} X$, where $U$ is 
	Uniform $(0,1)$ random variable, and $\overset{dist.}{=}$ represents equality 
	in distribution. Therefore, under Model (\ref{eq:qtwas2}),  the distribution of 
	gene 
	expression $X$  can be expressed as a convolution of a genotype-related 
	random variable $Z^\top \beta(U)$ and the 
	remainder term, $R(U)=\alpha_{0}(U)+C^\top\alpha(U)+ Q_\epsilon(U)$.
	Together with the assumption of independence between $X$ and the 
	covariates $C$, we  conclude that
	$$X_Z \overset{dist.}{=} Z^\top \beta(U),$$
	i.e. the genotype-related gene expression $X_Z$ has the same distribution as 
	$Z^\top \beta(U)$. Therefore, it is easy to derive that $\mathbb{E}(X_Z\mid 
	Z,X_Z\in A_k) = \int_{\tau_{k}}^{\tau_{k+1}} Z^\top \beta(u)du = 
	Z^\top\beta_{A_k},$ where 
	$\beta_{A_k} :=\int_{\tau_{k}}^{\tau_{k+1}}\beta(u)du $. Accordingly, the 
	estimator for $\beta_{A_k}$ is given by $\hat\beta_{A_k} = 
	\int_{\tau_k}^{\tau_{k+1}} \hat\beta(\tau) d\tau$, where $\tau_k$ and 
	$\tau_{k+1}$ define the range of $A_k$. Together with the estimated 
	$\hat{\beta}_{\rm GWAS}$ from the GWAS models, we can estimate 
	$\gamma_k$ by 
	\begin{eqnarray} \label{eq:estgamma}
		\hat\gamma_k
		&=& \frac{\widehat{cov}(Z^\top\beta_{A_k}, Z^\top
			\beta_{\rm GWAS})}{\widehat{var}(X_Z\mid X_Z\in 
			A_k)}=\frac{\hat\beta_{A_k}^\top  \widehat\Sigma_Z\hat\beta_{\rm 
				GWAS}}{\hat\sigma^2_{X_Z\in{A_k}}},
	\end{eqnarray}
	where $\hat\sigma^2_{X_Z\in{A_k}}$ is the variance of imputed gene 
	expression in the region $A_k$ and acts as an estimate of 
	$\sigma^2_{X_Z\in{A_k}}:=var(X_Z\mid X_Z\in A_k)$. Additionally, 
	$\widehat\Sigma_Z$ is the sample estimate of the covariance matrix of $Z$, 
	with its true value given by $$\Sigma_Z:=\text{var}(Z) = 
	\text{diag}\{\sigma_1, \cdots, \sigma_p\} D_Z \text{diag}\{\sigma_1, \cdots, 
	\sigma_p\},$$ where $\sigma_j$ stands for the standard deviation of the 
	$j$th SNP and $D_Z$ is the  
	linkage disequilibrium (LD),
	which describes the correlation structure of the genotypes $Z$. The LD 
	matrix, $D_Z$,  can be estimated using data from the GTEx 
	project or other external reference datasets.The standard deviation 
	$\sigma_j$ for each SNP can be estimated based on its minor allele 
	frequency.

	\paragraph{Constructing test statistics for the quantile-stratified  
	gene-trait association.}\label{sec:step2}
	The standard errors of $\hat\gamma_k$ can be estimated by
	$$se(\hat{\gamma}_k) = \sqrt{\frac{var(\epsilon_y)}{N_{\rm 
				GWAS}var(X_Z\mid X_Z \in 
			A_k)}}\approx\frac{\hat\sigma_{Y}}{\sqrt{N_{\rm 
				GWAS}}\hat\sigma_{X_Z\in{A_k}}},
	$$
	where $\hat\sigma_{Y}$ is the estimated standard deviation of trait $Y$, and 
	$N_{\rm GWAS}$ is the sample size of GWAS data. In the formula above, we 
	approximate $var(\epsilon_y)$ using $\hat\sigma_Y$, since the variance of 
	the trait explained by a single gene is typically minimal 
	\citep{hu2019statistical,o2017estimating}. 	
	
	
	The ratio  $\mathcal{Z}_k =\hat{\gamma}_k/se(\hat{\gamma}_k)$ naturally 
	forms a Wald-type test statistic for testing $H_0:\gamma_k = 0$ as below
	\begin{eqnarray}\label{eq:Zk}
		\mathcal{Z}_k = \frac{\hat{\gamma}_k}{se(\hat{\gamma}_{A_k})}&\approx& 
		\frac{\sqrt{N_{\rm GWAS}}}{\hat\sigma_Y\hat\sigma_{X_Z\in 
				A_k}}\hat\beta_{A_k}^\top 
		\widehat\Sigma_Z\hat{\beta}_{\rm GWAS},
	\end{eqnarray}
	where $\hat{\beta}_{\rm GWAS,j}$'s are estimated from separate marginal 
	linear regression models (eq. \eqref{eq:gwas}), and thus, they are correlated 
	due to the linkage disequilibrium (LD) structure among the SNPs $Z$. 
	Denote 	
	$\Delta = \frac{1}{\hat\sigma_{X_Z\in A_k}^2}\hat\beta_{A_k}^\top 
	\widehat\Sigma_Z\hat\beta_{A_k},$ then we have,
	under $H_0: \gamma_k =0$,
	$$\Delta^{-1/2}\mathcal{Z}_k \approx N\left(0, I\right),$$
	because $\frac{\sqrt{N_{\rm 
				GWAS}}}{\hat\sigma_Y}\Delta^{1/2}\hat\beta_{\rm GWAS} \approx 
				N(0,I)$ 
	under the null hypothesis of no SNP-trait associations 
	\citep{hu2019statistical}.

\begin{figure}[!ht]
		\centering
		\includegraphics[scale = 0.45]{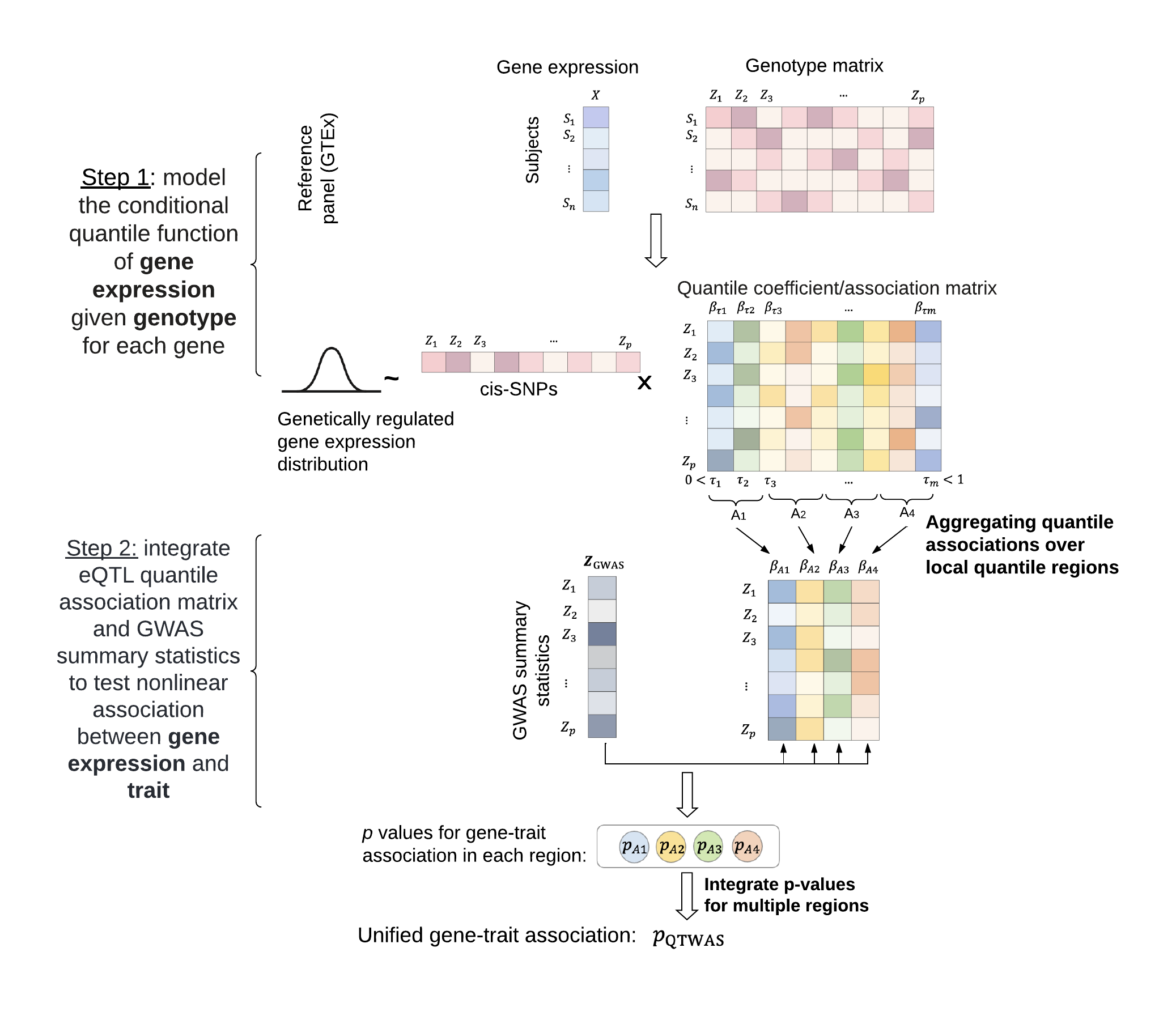}
		\vspace{-10mm}
		\caption{QTWAS flowchart with a specific region partition ($K=4$). }
		\label{fig:flow_chart}
	\end{figure}

	
	The $p$-value for testing $\gamma_k = 0$ in each region $A_k$ can be 
	computed as $p_k = 2\Phi(-|\Delta^{-1/2}\mathcal{Z}_k|)$, where 
	$\Phi(\cdot)$ is the standard normal CDF.  Finally, we combine all $p_k$'s 
	from $K$ regions by the Cauchy 
	combination method \citep{liu2020cauchy}, which offers a convenient 
	analytical solution to combine   $p$-values from  correlated tests. 
	Alternative methods for combining $p$-values could also be implemented, 
	such as Fisher's combination method or  the minimum $p$-value 
	\citep{dudoit2003multiple}. However, these methods often rely on the 
	assumption of independence among $p$-values (which is unrealistic in our 
	setting because of correlations across quantile regions) or require 
	computationally intensive approximations.  In Figure 
	\ref{fig:flow_chart}, we summarize the flowchart of QTWAS. We 
	provide further discussion on selecting $K$ and other implementation details 
	in 
	Section \ref{sec:implement}. 	
	
	\subsection{From linear models to generalized linear 
		models}\label{sec:binary}
	Although the derivation of the QTWAS test statistics above assumes a 
	linear GWAS model with a continuous $Y$, the same derivation 
	applies to generalized linear models with any link function. We denote the 
	GWAS data as 		$\{Y_i, \mathbf{Z}_i, \mathbf{C}_i\}_{i=1}^{N_{\rm 
			GWAS}}$, where $\mathbf{Z}_i = (Z_{i1},\dots, Z_{ip})$ represents 
	genotype data while  
	$\mathbf{C}_{i} = (C_{i1},\dots,C_{iq})$ represents covariates.  Assume 
	there is an arbitrary 
	link function $h(\cdot)$, such that $\mathbb{E}(Y_i \mid Z_{ij}, 
	\mathbf{C}_i) = h(\alpha_0+Z_{ij}\beta_{\rm GWAS, 
		j}+\mathbf{C}_i\eta)$ and $\mathbb{E}(Y_i\mid X_{Z,i}, \mathbf{C}_i) 
	= h(g_1(X_{Z,i})+g_2(\mathbf{C}_i))$. Common link functions include 
	$h(x) =\exp(x)/(1+\exp(x)) $ for logistic regression, $h(x) = x$ for linear 
	regression, $h(x) = \exp(x)$ for loglinear models. 
	
	We can treat the GWAS summary statistics from generalized linear 
	models as if they were estimated from linear regression with 
	pseudo-response $Y_i^*$. Specifically, we can define two models  
	\begin{eqnarray*}
		{\rm Model\ B:}&& \ \ Y_i^*= h^{-1}\{\mathbb{E}(Y_i\mid Z_{ij}, 
		\mathbf{C}_i)\}+\epsilon^*_i = \alpha_0+Z_{ij}\beta_{\rm GWAS, 
			j}+\mathbf{C}_i\eta+\epsilon^*_i,\\
		{\rm Model\ C:}&& \ \  Y_i^*= h^{-1}\{\mathbb{E}(Y_i\mid X_{Z,i}, 
		\mathbf{C}_i)\}+e^*_i = g_1(X_{Z,i})+g_2(\mathbf{C}_i)+e^*_i,
	\end{eqnarray*}
	where $\mathbb{E}(\epsilon^*_i) = \mathbb{E}(e^*_i) = 0$ and 
	$var(\epsilon^*_i)\approx var(e^*_i)\approx \sigma^2_{Y^*}$.

	A reasonable estimation for $\sigma_{Y^*}^2$ is $\sqrt{N_{\rm 
	GWAS}}\hat\sigma_jse(\hat\beta_{\rm GWAS, j})$, where 
	$\hat\sigma_j$ is the estimated standard deviation of the $j$-th SNP. This 
	estimate is
	based on the derivation of the standard error $se(\hat\beta_{\rm GWAS, j})$.   
	Therefore, the GWAS 
	summary statistics $\{\hat\beta_{\rm GWAS,j}, se(\hat\beta_{\rm GWAS, j})\}$, 
	which 
	were obtained from  generalized linear models using $\{Y_i, Z_{ij}, 
	\mathbf{C}_i\}_{i=1}^{N_{\rm GWAS}}$,  can be equivalently viewed as if they 
	were estimated from 
	linear models with the pseudo-response data $\{Y^*_i, 
	Z_{ij}, \mathbf{C}_i\}_{i=1}^{N_{\rm GWAS}}$. As a result, the derivation 
	of the QTWAS statistics remains valid, meaning that the GWAS summary 
	statistics can be used in eq. \eqref{eq:Zk} regardless of the specific models 
	from which they were originally generated.
	
	\subsection{Model implementation}\label{sec:implement}
	
	\subsubsection{Implementation details in the GTEx data}
	We  trained the gene expression prediction model for 49 tissues from the 
	GTEx project (v8), as described below. Gene expression levels were 
	normalized 
	and adjusted for covariates and confounders, including sex, sequencing 
	platform, and the top five principal components of genotype data,  as well as 
	the top 15 
	probabilistic estimation of expression residuals (PEER) factors 
	\citep{hu2019statistical, stegle2010bayesian}. We  considered 
	protein-coding genes, removed ambiguously stranded SNPs, and only 
	considered ref/alt 
	pairs A/T, C/G, T/A, and G/C. SNPs with minor allele frequency less than 0.01 
	were excluded from the analyses. For each gene, we used SNPs between 
	1Mb 
	upstream and downstream of the transcription start site. The LD matrix 
	$D_Z$ is estimated from the genotype data in the GTEx data. 
	
	\subsubsection{Variant screening procedure}\label{subsec:vs} Based 
	on empirical evidence, $X_Z$ often depends on a sparse set of SNPs 
	\citep{barbeira2018exploring, 
		gamazon2015gene}. Most existing TWAS approaches use penalized 
	linear regression to select significant SNPs associated with the mean of 
	$X_Z$ 
	\citep{barbeira2018exploring,gamazon2015gene,hu2019statistical}, which 
	may not be optimal at identifying more local (quantile-stratified) associations. 
	We introduce a new variant screening procedure based on the quantile rank 
	score test to identify important SNPs separately for each region $A_k$. 
	Specifically, 
	we aggregate multiple quantile rank score tests 
	\citep{gutenbrunner1993tests} 
	at selected quantile levels within $A_{X,k}$ to select region-specific SNPs 
	while controlling the false discovery rate at the $5\%$ level using the method 
	of 
	\cite{benjamini1995controlling}. The new screening procedure is more 
	effective at identifying heterogeneous distributional associations and 
	non-gaussian 
	errors. We outline the detailed algorithm and a flowchart of the screening 
	procedure in 
	Appendix Section 1.1.  Note that the variant screening step is a data 
	pre-processing procedure for training the 
	Genotype-GeneExpression model by GTEx data, which is independent of 
	the subsequent steps for constructing test statistics. Thus, it does not affect 
	the multiple testing burden at the gene-level QTWAS $p$-values.
	
	\subsubsection{Selection of $K$}
	The length of $A_k$ and the number of regions ($K$) can be set 
	depending on applications. Though the piecewise linear approximation eq. 
	\eqref{eq:linear_approx} assumes $\cup_{k=1}^K A_k = (0,1)$ for 
	estimation, this assumption can be relaxed in the context of hypothesis 
	testing. To test the overall 
	associations between $X_Z$ and $Y$, integrating $\beta(\tau)$'s over a 
	larger region 
	$A_k$ is especially helpful in detecting weak genetic associations. Based on 
	our 
	empirical experience, $K = 3$ or $4$ should be sufficient to detect 
	homogeneous 
	associations such as location shift, and a relatively larger $K$ (e.g., 
	$K=9$) 
	facilitates detecting local associations, whereas a very large $K$ is not 
	recommended because of the risk of power loss.  As the underlying 
	association patterns are unknown in real applications, one can further 
	consider using 
	partially overlapped regions and multiple choices of $K$ to improve the 
	power. In 
	practice, we use the Cauchy combination method to combine results from 
	$K\in\{3,4, 5, 9\}$ with slightly overlapped regions to obtain robust and 
	powerful results 
	(see the region segments in Appendix Section 1.2). This approach is 
	data-driven, 
	insensitive to the underlying association patterns, and avoids selecting $K$ 
	as a tuning parameter. Furthermore, for a specified $K$, we recommend 
	considering the partition such that $\cup_{k=1}^K A_k $ covers the 5\% 
	percentile to 
	the 95\% percentile of the value of $X_Z$. We do not recommend 
	considering 
	$\tau<0.05$ and $\tau>0.95$, as coefficients of extreme quantiles are more 
	challenging 
	to estimate. Note that excluding the extreme regions $\tau\in(0,0.05)$ and 
	$\tau\in(0.95,1)$ may lead to loss of power if local association only manifest 
	at these extreme 
	tails. Investigators can also choose the regions based on specific applications.
	
	\section{Simulation studies}\label{sec:simu}
	\subsection{Simulation settings}\label{sec:simu_settings}
	The simulation studies are based on the data in whole blood tissue from 
	GTEx v8 ($n=670$). We generate gene expression based on the genotype 
	data on 670 
	individuals from GTEx (see details below on the genotype-gene expression 
	models). Then, we resample $n=1,000$ subjects and generate their trait 
	values based on their genotypes. For each gene, $Z$ includes all SNPs 
	within $\pm 1$ Mb from its TSS. Gene expression $X$ is normalized before 
	analysis as common practice in genetic association tests. The set of 
	covariates $C$ includes the top five principal components, top 15 PEER 
	factors, platform, and sex. Similar to \cite{hu2019statistical}, we randomly 
	select 500 genes and generate the gene expression data and traits 
	independently for each gene, as described below. 
	
	\medskip
	
	To evaluate Type I error, we generate the gene expression $X$  from the 
	model: 
	$X = Z^\top \beta+ C^\top \alpha_x+\epsilon_x$, in which $\beta$ is 
	estimated based on true GTEx data via the elastic net with the tuning 
	parameter set 
	as $0.5$. The trait $Y$ is generated by $Y = C^\top \eta+ e$. Both error 
	terms $e$ and $\epsilon_x$ follow a standard normal distribution;  
	$\alpha_x$ and 
	$\eta$ are vectors with each element randomly drawn from ${\rm Unif}(0,1)$. 
	This 
	null model preserves the associations between gene expression and SNPs 
	from 
	GTEx data but assumes no gene-trait association. A similar setting has been 
	simulated in \cite{hu2019statistical}. 
	
	\medskip
	
	For power analyses, we consider three different 
	Genotype-GeneExpression models, and we assume a  simple linear 
	Genotype-Trait model to mimic the setting of GWAS summary statistics 
	from linear models. 
	
	\medskip
	
	\underline{Genotype-GeneExpression models.} We consider the following 
	three models: (a) Location shift: $X =  Z^\top \beta+C^\top\alpha_x + 
	\epsilon_x$; (b) Location-scale: $X =  Z^\top \beta +C^\top\alpha_x+ 
	(1+0.5 Z^\top \beta)\epsilon_x$; (c) Local  signal: $Q_{X}(\tau>0.7 \mid Z, 
	C) = 5\frac{\tau-0.7}{1-0.7}Z^\top 
	\beta+C^\top\alpha_x+F^{-1}_{\epsilon_x}(\tau)$.
	
	In the location shift model (a), genetic variants $Z$ only affect the 
	mean of $X$, while in the location-scale model (b), genetic variants $Z$ 
	affect 
	both the mean and variance of $X$. In the local signal model (c), variants $Z$ 
	only affect part of the distribution of $X$, i.e., $Z$ only affects the upper 
	quantile ($>0.7$th quantile) of $X$. In each of the three scenarios, we 
	consider 
	two error distributions for $\epsilon_x$: standard normal and Cauchy 
	distributions, 
	where the Cauchy distribution is commonly considered as a challenging 
	case of heavy-tailed distribution in association studies 
	\citep{song2017qrank,wang2022integrated}. Under models (b) and (c), 
	when the quantile specific coefficients are different across quantiles, 
	the Genotype-GeneExpression association is heterogeneous, and the 
	transcriptome-wide association is nonlinear  (Appendix Section 
	2.1). 
	
	\medskip
	
	\underline{Genotype-Trait model.} We consider a simple linear model $Y =  
	Z^\top \beta_{\rm GWAS}+ C^\top \eta + e$, where $e$ follows a standard 
	normal distribution.

	To illustrate the performance in different scenarios, we randomly select  
	1\% of SNPs from the 2Mb region around TSS to be causal (i.e., with 
	non-zero effect sizes $\beta$ and $\beta_{\rm GWAS}$).
	We set $\beta_{\rm GWAS}=\mathbf{1}_p$ and $\beta = 2\cdot 
	\mathbf{1}_p$ for local signal model,  $\beta_{\rm GWAS}= 
	0.2\cdot\mathbf{1}_p$ and  $\beta = 0.4\cdot\mathbf{1}_p$ for 
	location-scale model, and  $\beta_{\rm GWAS}=0.1\cdot\mathbf{1}_p$ and 
	$\beta = 0.2\cdot\mathbf{1}_p$  for location shift model, where 
	$\mathbf{1}_p$ represents a column vector with all elements being 1. 
	$\alpha_x$ and $\eta$ are vectors with each element randomly drawn from 
	${\rm Unif}(0,1)$.

	For power analyses, we repeat the data generation procedure two times 
	per gene and report the statistical power based on 1,000 replicates at the 
	significance threshold $\alpha = 2.5e$-6 (corresponding to the usual 
	Bonferroni threshold when testing 20,000 protein-coding genes). For type I 
	error 
	analysis, we repeat the procedure for each gene 20,000 times and report the 
	results 
	based on $10^7$ replicates at different significance thresholds ranging 
	from 0.05 to $2.5e-6$. In addition, we compare the proposed framework with 
	S-PrediXcan (note that we have re-implemented 
	S-PrediXcan as it needs to be trained based on simulated data, and we 
	denote it as ``S-PrediXcan$^*$"). For our method, we report results based 
	on a 
	single 		region partition ($K = 3/4/5/9$) and the unified results combining all 
	partitions. 
	The detailed partitions are described in Appendix Section 1.2. For one 
	gene, we generate random $p$-values $p\sim{\rm Unif}(0,1)$ if the elastic 
	net model in 
	S-PrediXcan$^*$ does not select any variables, or if none of the four 
	regions 
	$A_1$-$A_4$ in QTWAS has valid $p$-value (e.g., no variant is selected).
	
	\medskip
	
	\subsection{Simulation results}\label{sec:simu_results}
\begin{figure}
		\centering
		\includegraphics[scale = 
		0.28]{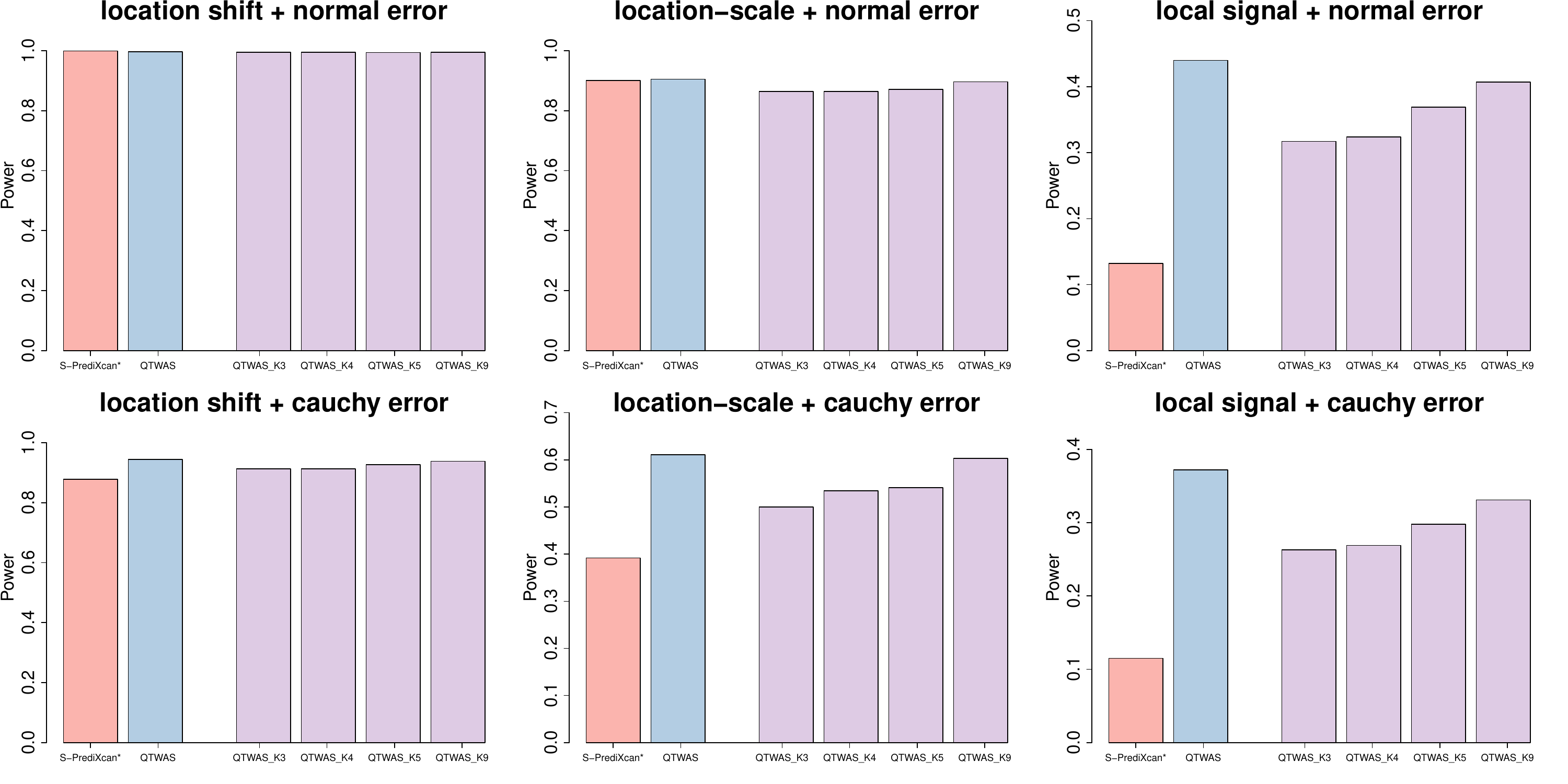}
		\caption{Power of S-PrediXcan$^*$ and QTWAS under alternative 
			models.  
			The significance threshold is $\alpha = 2.5e-6$. ``QTWAS" 
			combines the $p$-values of $K=3/4/5/9$ via the Cauchy combination 
			method. }
		\label{fig:power}
	\end{figure}

	The type I error for QTWAS, either with a single choice of $K$ or a unified 
	result based on four choices of $K$, is controlled at all significance levels 
	(Table \ref{tab:H0_new}). Regarding power performance, QTWAS, 
	combining different quantile intervals, has improved  power in most 
	scenarios compared to S-PrediXcan$^*$ (Figure \ref{fig:power}).  When 
	the error is Gaussian, both methods have comparable power for the 
	location shift and location-scale models. However, QTWAS has 
	substantially improved power over S-PrediXcan$^*$ when the association 
	is local and only at upper quantiles. When the error follows the Cauchy 
	distribution, QTWAS performs well compared to S-PrediXcan. Additionally, 
	we observe that QTWAS is not very sensitive to the choice of $K$, and the 
	unified approach performs best.

	\begin{table}[ht]
		\centering
		\begin{tabular}{r|r|rrrrr}
			\hline
			$\alpha$&S-PrediXcan*&\multicolumn{5}{|c}{\underline{QTWAS}} \\
			&  &Unified&$K=3$&$K=4$&$K=5$&$K=9$ \\ \hline
			0.05 & 5.024E-02 & 5.154E-02 & 4.993E-02 & 4.910E-02 & 
			4.924E-02 
			& 
			4.937E-02 \\ 
			1e-2 &  1.075E-02 & 9.784E-03 & 9.908E-03 & 1.017E-02 & 
			9.841E-03 
			& 
			9.843E-03 \\ 
			1e-3 & 1.248E-03 & 6.536E-04 & 9.267E-04 & 8.847E-04 & 
			1.004E-03 
			& 
			8.584E-04 \\ 
			1e-4 &  9.770E-05 & 3.900E-05 & 2.380E-05 & 6.140E-05 & 
			6.490E-05 
			& 
			5.910E-05 \\ 
			1e-5 &  4.600E-06 & 1.800E-06 & 1.700E-06 & 1.200E-06 & 
			1.300E-06 
			& 
			1.500E-06 \\ 
			2.5e-6 &1.400E-06 & 1.200E-06 & 2.000E-07 & 3.000E-07 & 
			3.000E-07 & 1.200E-06 \\
			\hline
		\end{tabular}
		\caption{Type I error results for  S-PrediXcan$^*$ and QTWAS ($n_{\rm 
				GTEx} = 670$), as well as for quantile region stratified QTWAS 
			based on 
			$10^7$ replicates. ``Unified" combines the $p$-value of $K=3/4/5/9$ 
			via 
			the Cauchy combination method. }\label{tab:H0_new}
	\end{table}

	We further illustrate the power improvement of QTWAS using the partition 
	$K=4$ as an example. We provide the figure of region-specific power of 
	the QTWAS test statistics under three models in Appendix Section 2.1. The 
	power of QTWAS in each region reveals the true underlying signal. For 
	example, 
	the location shift model with normal errors has equally high power in each 
	region, and the location-scale model with normal errors shows increasing 
	power 
	from lower quantile to upper quantile, corresponding to the assumptions of 
	our 
	model. For local signal models, we observed a dominant power boost for 
	QTWAS, owing to the power of QTWAS test statistics in the upper quantile 
	region ($A_4$), corresponding to the true signals being simulated at upper 
	quantiles (i.e., $\tau>0.7$). Therefore, the region-specific quantile test 
	statistics can reveal more complex and detailed association patterns. 
	
	\medskip
	
	Under alternative models, we assess the robustness of the QTWAS 
	approach. That is, we report how sensitive the QTWAS is to the choice of 
	$K$. Among 
	the significant results of QTWAS in Figure \ref{fig:power}, we report the 
	proportion 	identified by at least two partitions with the significance threshold 
	$2.5e-6$ (see Appendix Section 2.2). We observe that most of the QTWAS 
	discoveries are identified by at least two partitions for all models, with the 
	proportion slightly decreasing for the models with a higher level of 
	heterogeneity. Overall, these results suggest that QTWAS results are robust.

	\subsection{Model evaluation}\label{sec:model_eval}
	\subsubsection{Imputation accuracy} 
\begin{figure}[!ht]
		\centering
		\includegraphics[scale = 0.5]{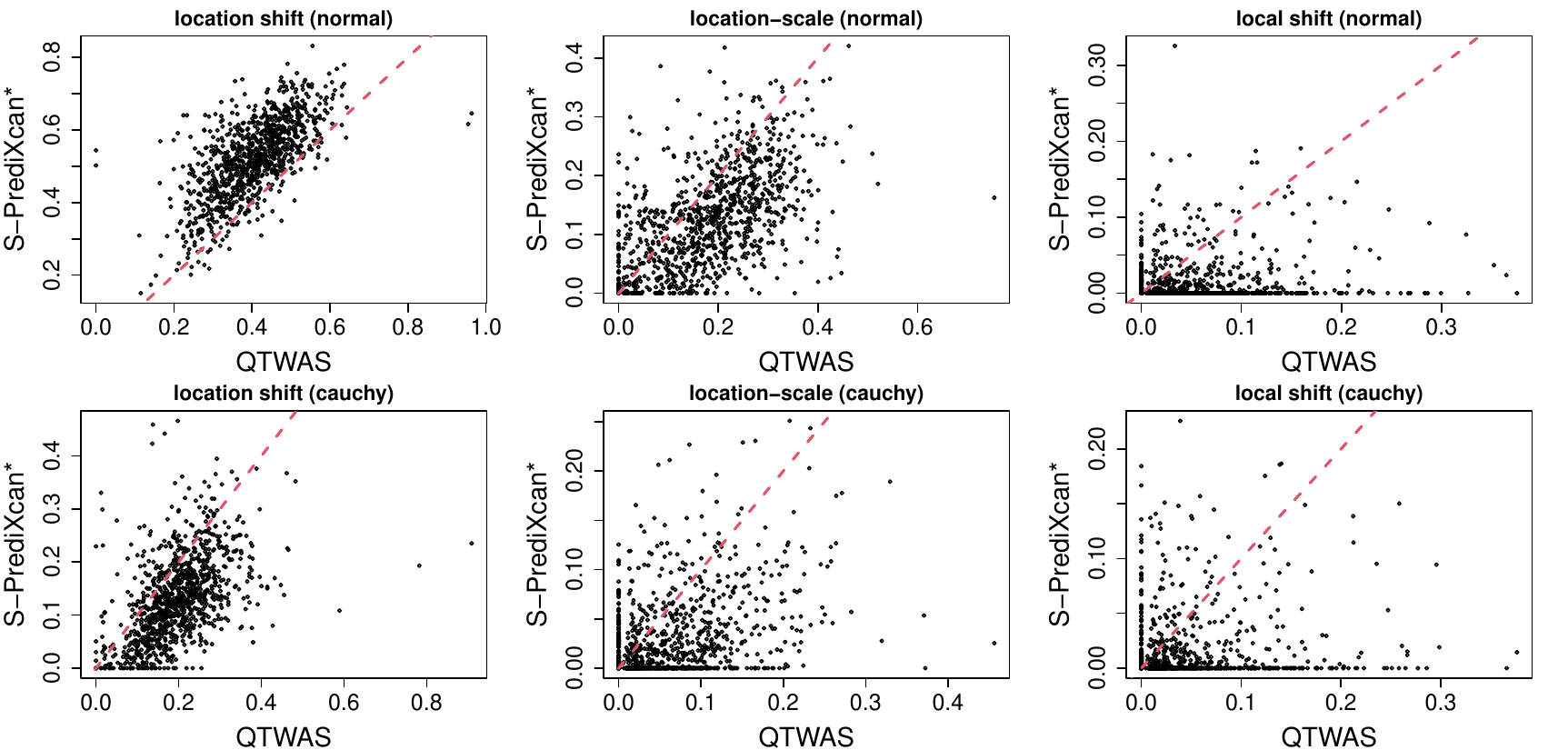}   
		\caption{Model explained deviance for QTWAS and S-PrediXcan$^*$. 
			We 	consider the location shift, location-scale, and local signal models 
			with 
			normal errors and Cauchy errors. }
		\label{fig:R}
	\end{figure}

To evaluate the accuracy of the 
	gene expression imputation model \eqref{eq:rq_model_method}, we 
	consider the goodness of fit criterion $R^Q(\tau) = 1-\hat{V}(\tau)/\tilde 
	V(\tau)$ \citep{koenker1999goodness}, a measure of explained 
	deviance by the quantile model associated to genetic effects at a fixed 
	quantile level, where $\hat V(\tau) = \min\sum_i^n \rho_\tau(X_i-C_i^\top 
	\alpha_\tau-Z_{i}^\top\beta_\tau - \alpha_{0,\tau})$ and $\tilde V(\tau) = 
	\min\sum_i^n \rho_\tau(X_i-C_i^\top \alpha_\tau - \alpha_{0,\tau})$ are 
	optimized  quantile loss under the alternative and null model, respectively. 
	It is a natural analog to $R^2$ in linear models. We use $K=4$ as an 
	example and consider the largest $R^Q(\tau)$ over the four intervals as 
	the explained deviance by QTWAS, which coincides with the fact that the 
	Cauchy combination is practically driven by the smallest $p$-value in the 
	combination. To compare the imputation accuracy for QTWAS and 
	S-PrediXcan$^*$, we plotted $R^Q$ against $R^2$ (Figure \ref{fig:R}). 
	Except for the location shift model with normal error, QTWAS generally 
	explained more deviance than S-PrediXcan$^*$. Specifically, in the 
	location-scale and local signal models, S-PrediXcan$^*$ explains a low 
	proportion of the total deviance, indicating a relatively poor goodness of fit 
	compared to quantile models.
	
	\subsubsection{Evaluation of variant screening procedure}  To evaluate 
	the quantile variant screening, we measure the canonical correlation 
	between selected sets and the causal set in the three alternative models 
	with normal or Cauchy errors. For ease of presentation, we again consider 
	$K=4$ as an example and use ``QTWAS$_{A_k}$" to denote the results 
	for region $k=1, 2, 3, 4$. The proportion of replicates with a correlation 
	greater than 0.95 is reported in Table \ref{tab:ccor} based on 1,000 
	replicates. For the location shift model, both QTWAS and S-PrediXcan$^*$ 
	select 
	SNP sets highly correlated with the true causal set. In the location-scale 
	model, the proportion of replicates selecting highly correlated SNPs is 
	increasing with the quantile levels and is comparable to S-PrediXcan$^*$ 
	for interval $A_4$, consistent with the power results. In the local  shift 
	model, QTWAS selected a set of SNPs with high correlation with the true 
	causal set  in the upper quantile interval $A_4$ more often than 
	S-PrediXcan$^*$, as expected. The ability of QTWAS to select variants 
	that are more correlated to the underlying causal variants in heterogeneous 
	cases is likely due to the specific quantile-oriented screening procedure 
	we use here.

	\begin{table}[!ht]
		\centering
		\begin{tabular}{l|rrrrr}
			\hline
			Model   &   QTWAS$_{A1}$ & QTWAS$_{A2}$ & QTWAS$_{A3}$ & 
			QTWAS$_{A4}$&S-PrediXcan$^*$\\ \hline
			location shift&         97.9\%& 99.1\%& 99.1\%& 99.0\%& 99.8\%\\
			location-scale &31.3\% &54.2\% &70.4\% &83.7\% &85.4\%\\
			local  shift&4.5\%& 0.5\%& 0.5\%& 39.1\%& 15.4\% \\
			
			\hline
		\end{tabular}
		\caption{The proportion of replicates (out of 1,000) with canonical 
			correlation values between the selected variable set and the true 
			causal set 
			greater than 0.95.}
		\label{tab:ccor}
	\end{table}

		\subsection{Comparisons to sMiST}\label{sec:sMiST} 
		We also compare the performance of S-PrediXcan and QTWAS to sMiST 
		\citep{dong2020general}, a method based on mixed effect models to 
		test the total effect of genetic variants, including their direct effects and 
		indirect 
		effects through gene expression. Though sMiST can also be performed 
		based on summary statistics, the goal of 
		sMiST is	to test not only the effect of imputed gene expression (e.g., 
		similar 
		to TWAS) but also the direct effect of genetic variants, which violates the 
		second assumption outlined in Section \ref{subsec:assu} for classic 
		TWAS. By adding the genotype information into the 
		Genotype-Trait model with random effects, their model can be written in 
		our notations as below:
		\[
		g\{E(Y\mid X, Z, C)\} = C^\top\eta + X\gamma_x + Z^\top\delta, 
		\]
		where $\delta_j$ for $j=1,\cdots, p$ are random effects with mean zero 
		and variance $\sigma^2_\delta$. The null hypothesis that sMist tests is 
		$H_0: 
		\gamma_x = 0$ and $\sigma^2_\delta = 0$. Though S-PrediXcan and 
		sMiST are both based on GWAS summary 
		statistics, their performance are not directly comparable because sMiST 
		needs S-PrediXcan estimated coefficients as an input, and it tests both 
		fixed effects and random effects. Thus, it is more powerful than 
		S-PrediXcan when the $\sigma^2_\delta \neq 0$. Nevertheless, we 
		applied sMiST to the three models we considered in our simulations. The 
		\texttt{R} package is 
		obtained from the author's website 
		(https://research.fredhutch.org/hsu/en/software.html). We report 
		results based on three $p$-value combination procedures offered by 
		their 
		\texttt{R} package. From Table \ref{tab:sMiST}, we can see that sMiST is 
		more powerful than S-PrediXcan in the local model, as expected. 
		Compared to QTWAS, sMiST is comparable in the 
		location and location-scale models but less powerful in the local model, 
		which suggests that it is promising to develop a similar mixed effects 
		model under the quantile framework to further improve the power of the 
		association test. 
		
		\begin{table}[!ht]
			\centering
			\begin{tabular}{r|r|r|rrr}
				\hline
				Setting& S-PrediXcan* & QTWAS & \multicolumn{3}{c}{sMiST} \\
				
				& & & p.oMiST&p.aMiST&p.fMiST\\ 
				\hline
				
				location shift & 0.999 & 0.996 & 0.997 & 0.996 & 0.997  \\ 
				location-scale& 0.901 & 0.905  & 0.911 & 0.909 & 0.911 \\ 
				local shift&0.132 & 0.440  & 0.244 & 0.237 & 0.236 \\ 
				\hline
			\end{tabular}
			\caption{Comparisons with sMiST. Power results based on 1000 Monte 
				Carlo 
				replicates.}\label{tab:sMiST}
		\end{table}

	\subsection{Additional simulations}
	The previously presented local signal model is in favor of quantile 
	regression, as 
	the association only appears at upper quantiles. In the Appendix (Section 
	3.4), 
	we consider two additional models: $Q_X(\tau\mid Z, C) = 
	Z^\top\beta(\tau)+C^\top\alpha_x+F^{-1}_{\epsilon_x}(\tau)$ with 
	$\beta(\tau) 
	= \sqrt{\tau}$ and $\beta(\tau) = \sin(2\pi \tau)$, respectively.
	Different from the previous local model, in which $\beta(\tau)\neq 0$ only 
	for 
	$\tau \in (0.7, 1)$, $\beta(\tau)$ here changes smoothly for $\tau\in(0,1)$. 
	For 
	these two heterogeneous models, QTWAS also outperformed 
	S-PrediXcan*. 
	Specifically, QTWAS is powerful for the model with $\beta(\tau) = 
	\sin(2\pi\tau)$ 
	while S-PrediXcan* has almost no power because there is no association 
	when $\tau=0.5$. We further consider the location model with unobserved 
	gene-environment interaction for the Genotype-GeneExpression model 
	(Appendix Section 3.3). Results suggest that when there exists 
	gene-environment interaction, but the environmental factor is unobserved, 
	QTWAS is equivalently powerful or more powerful than S-PrediXcan in 
	detecting gene-trait associations. 
	
	\medskip

		We further conduct simulations based on other tissues from GTEx data 
		with 
		smaller sample sizes. In particular,  we considered the breast mammary 
		tissue ($n=396$) and the brain cerebellum tissue ($n=209$). The 
		Genoytpe-GeneExpression model and the Genotype-Trait model remain 
		the same as in Section \ref{sec:simu_settings}. With a smaller sample 
		size of the genotype-gene expression data, we still observe significant 
		power 
		improvement of QTWAS over S-PrediXcan*. The detailed results are shown 
		in 
		Appendix Section 3.5.

	Additionally, we consider another data-generating 
		mechanism under the same Genotype-GeneExpression model and the 
		Genotype-Trait model, but with cis-heritability controlled at 10\% and 
		30\% levels, respectively. At both levels, QTWAS exhibits higher power 
		than S-PrediXcan. Results are shown in Appendix Section 3.6. 
 At the cis-heritability levels 10\% 
		and 30\%, we also consider the scenario with horizontal pleiotropy. That 
		is, we 
		add a direct genetic effect in the Genotype-Trait model in addition to the 
		classic 
		homogeneous model (Appendix Section 3.7). The 
		Genotype-GeneExpression model is the same as before. Results 
		suggest that QTWAS is more powerful than S-PrediXcan at 10\% and 
		30\% 
		cis-heritability levels. However, we acknowledge that with horizontal 
		pleiotropy, 
		both 
		S-PrediXcan and QTWAS cannot infer causality and only detect 
		associations (see Appendix Figure 8); a detailed discussion of robust 
		methods that detect causality against horizontal pleiotropy is provided in 
		Section \ref{sec:discussion}.

	\section{Applications}\label{sec:LDL}
	
	We apply S-PrediXcan and QTWAS to publicly available GWAS 
	summary statistics from UK Biobank (UKBB) from \cite{panukbb}. 
	 Specifically, we focus on two traits, 
		one continuous (low-density lipoprotein (LDL)) and a binary trait
		(schizophrenia (SCZ)) \citep{pardinas2018common}. For LDL  (``LDL 
		direct, 
		adjusted by medication''), $N_{\rm GWAS}=398,414$ and we leverage 
		data on whole-blood tissue from GTEx with $n=670$. For SCZ, we 
		leverage 
		summary statistics on $N_{\rm GWAS}=35,802$ individuals and gene 
		expression 
		data on 13 brain tissues from 
		GTEx. For QTWAS, we consider different quantile partitions with $K \in 
	\{3,4,5,9\}$ and use the Cauchy combination to combine all $p$-values.
	Note that we only keep genes with  $R^Q>0.1$ for QTWAS, and, similarly, we 
	consider $R^2>0.1$ for 
	S-PrediXcan results. 
	 We focus on protein-coding genes and further restrict to the 
		set of 6,560  genes with valid pre-trained S-PrediXcan models available 
		from the PredictDB website \citep{SPrediXcan}. Further, we use genomic 
		control \citep{devlin1999genomic} to adjust for possible inflation induced 
		by polygenic effects, although future model developments based on mixed 
		effect models will be implemented in the QTWAS framework (see Section 
		\ref{sec:discussion}). 
		The significance threshold we used is $2.5e-6$.

			\subsection{Results for LDL}
\begin{figure}[!ht]
				\centering
				\includegraphics[width=0.49\linewidth]{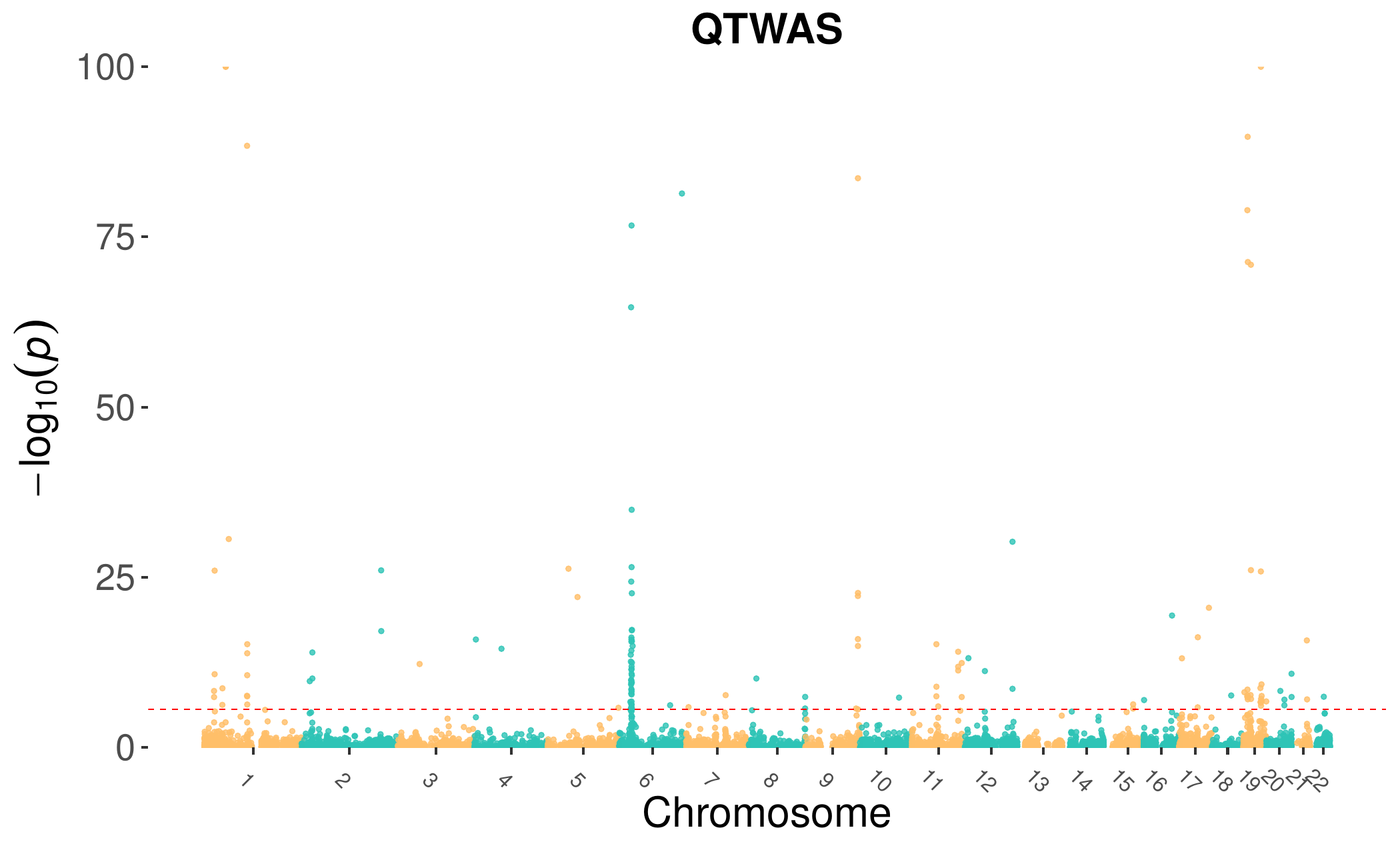}
				\includegraphics[width=0.49\linewidth]{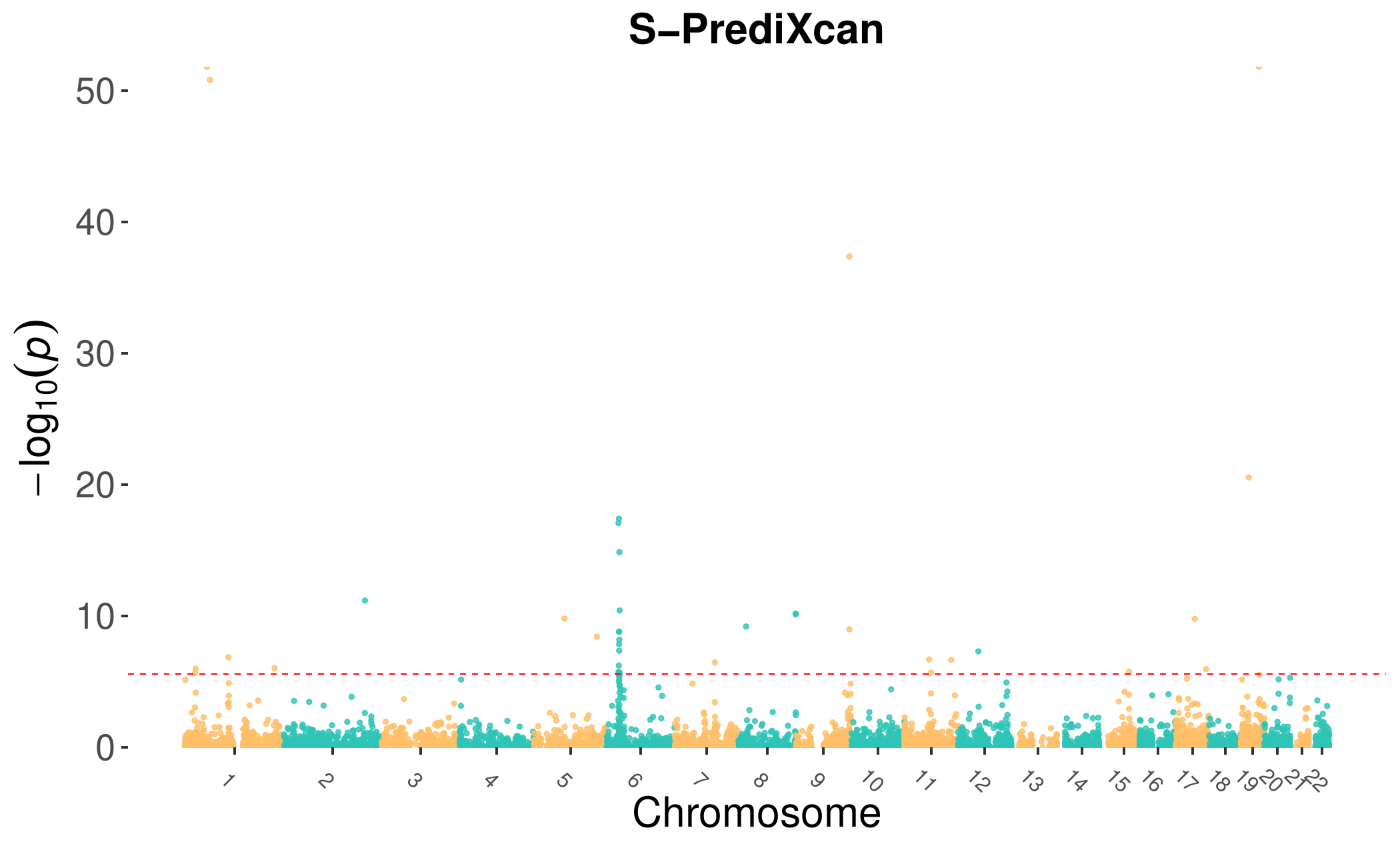}
				\caption{Manhattan plot for LDL with significance threshold 
					2.5$e-6$.}
				\label{fig:manhattan_LDL}
			\end{figure}

			QTWAS identified 136 genes while S-PrediXcan 
			identified 39 genes, with 29 genes identified by both methods (Figure 
			\ref{fig:manhattan_LDL}).
			\medskip
			
			Further, we explore the reproducibility of our findings using another 
			GWAS of lipid levels \cite{graham2021power}, which contains 1.65 million 
			individuals of mixed ancestries (although 
			the European ancestry is dominant at 79.8\%). Among the 107 genes that 
			were uniquely 
			identified by QTWAS in UKBB, 73 genes were also 
			significantly identified in the new study at significance level $2.5e-6$. 
			In 	contrast, for the 10 genes that were uniquely identified by 
			S-PrediXcan, none of 
			them were successfully reproduced in the second study. 
			These results suggest that QTWAS may be more powerful.

		\subsection{Assessing nonlinearity and functional enrichment 
			analysis}
		As the two-step TWAS framework combines both eQTL and GWAS 
		information from two datasets, the genes only identified by QTWAS but 
		not  by S-PrediXcan are highly likely to be detected due to nonlinear 
		associations. 
		To further assess the heterogeneity and nonlinearity of 
			gene expression for the genes uniquely identified by QTWAS, we first 
			obtain 
			their residuals from an elastic net model that fits gene expression to 
			genotypes and covariates. We then fit another elastic net model with 
			the squared residuals against the genotypes. The explained deviations in 
			this model, similar to $R^2$ in linear models, measure the correlation 
			between the squared residuals and genotypes. A higher value indicates
			stronger heteroskedasticity, as the squared residuals represent the 
			variance in 
			gene expression. We compare the nonlinearity between the group of 
			genes 
			identified by both QTWAS and S-PrediXcan and those identified only by 
			QTWAS. 
			Our results show that the genes identified exclusively by QTWAS exhibit 
			a 
			higher degree of nonlinearity (Table \ref{tab:res_r2}).

		\begin{table}
			\centering
			\resizebox{\textwidth}{!}{
			\begin{tabular}{llll}
				\hline
				variance of gene expression&number of genes  & only identified  
				& 
				identified by both \\ 
				explained by genotypes &identified by QTWAS &  by QTWAS &   
				QTWAS and S-PrediXcan\\ 
				\hline
				less than 5\% & 96& 78\% & 22\% \\ 
				between 5\% and 10\% & 24& 71\% & 29\% \\ 
				greater than 10\% & 16&94\% & 6\% \\ 
				\hline
			\end{tabular}
		}
			\caption{Analysis of heterogeneity and nonlinearity of genes 
				identified by QTWAS using GTEx data. }\label{tab:res_r2}
		\end{table}

		We further performed 
		enrichment analysis to explore the function of the 107 genes uniquely 
		identified by QTWAS but not by S-PrediXcan. We used the ToppGene 
		database (https://toppgene.cchmc.org/enrichment.jsp) and focused on 
		traits enriched in this group of genes. As presented in Table 
		\ref{tab:enrich}, the most enriched trait is LDL measurement, which is 
		our primary response trait. Following LDL measurement, the traits with the 
		highest number of gene hits are total cholesterol, apolipoprotein B, and 
		triglycerides, all of which are closely associated with LDL levels. The 
		fifth trait, cholesteryl ester, is a form of cholesterol in which a fatty acid 
		is attached to the cholesterol molecule. In total, 61 
		out of the 107 genes are enriched across these five traits.

		\begin{table}[!ht]
			\centering
			\resizebox{\textwidth}{!}{
			\begin{tabular}{lcc}
				\hline
				Trait& q-value (Bonferroni)& Number of Gene Hits\\ 
				\hline
				low density lipoprotein cholesterol measurement & 5.932e-32 &  
				52 \\ 
				total cholesterol measurement & 1.709e-24 &  43 \\ 
				apolipoprotein B measurement & 2.034e-19 & 28\\ 
				triglycerides measurement & 5.628e-14 &  36 \\ 
				cholesteryl ester measurement 
				& 1.758e-12 &  16 \\ 
				
				\hline
			\end{tabular}
		}
			\caption{Genes uniquely identified by QTWAS are enriched in 
			LDL-related traits in ToppGene.}
			\label{tab:enrich}
		\end{table}

\subsection{Application to 
	schizophrenia}
\begin{figure}[!ht]
		\centering
		\includegraphics[width=0.49\linewidth]{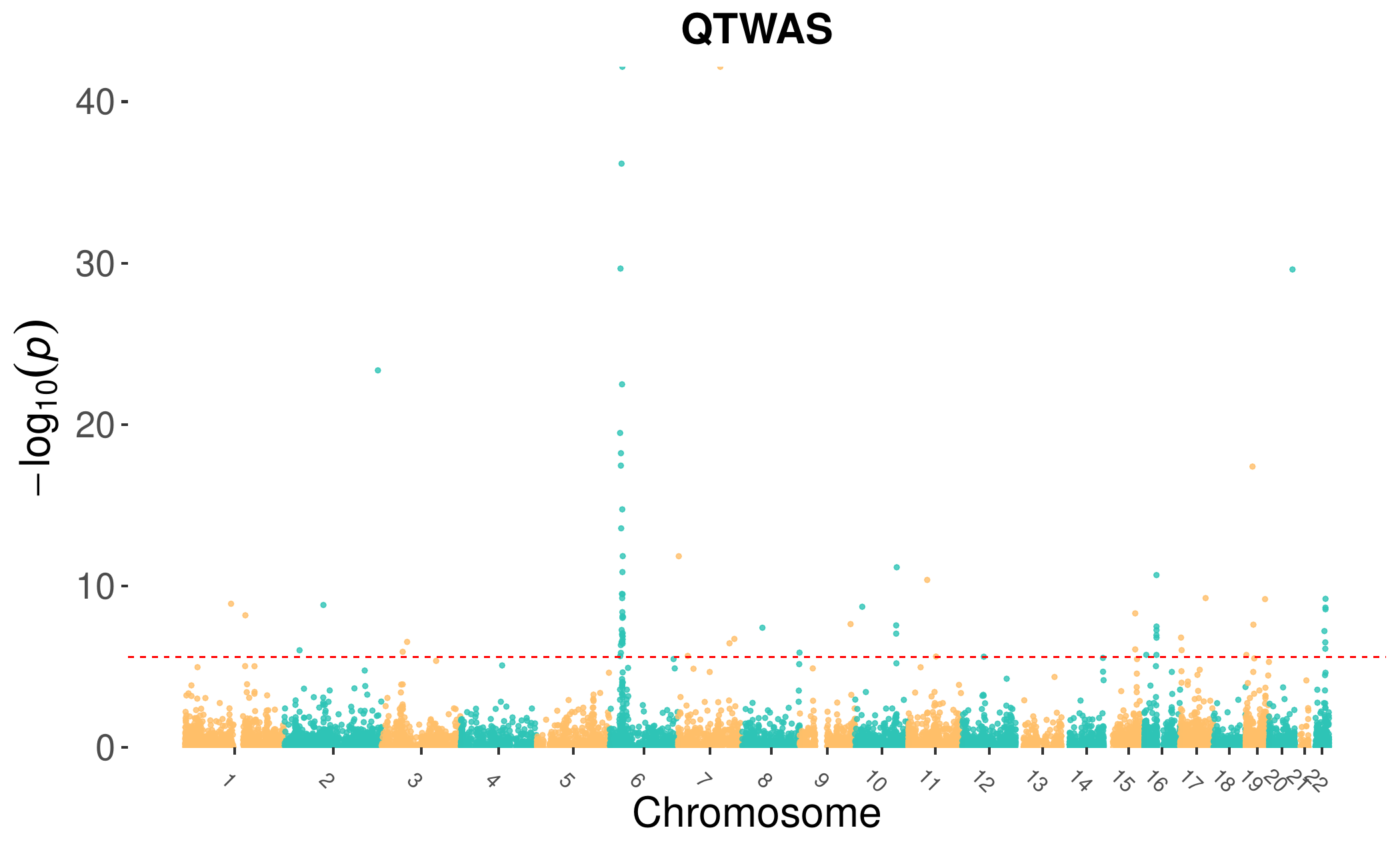}
		\includegraphics[width=0.49\linewidth]{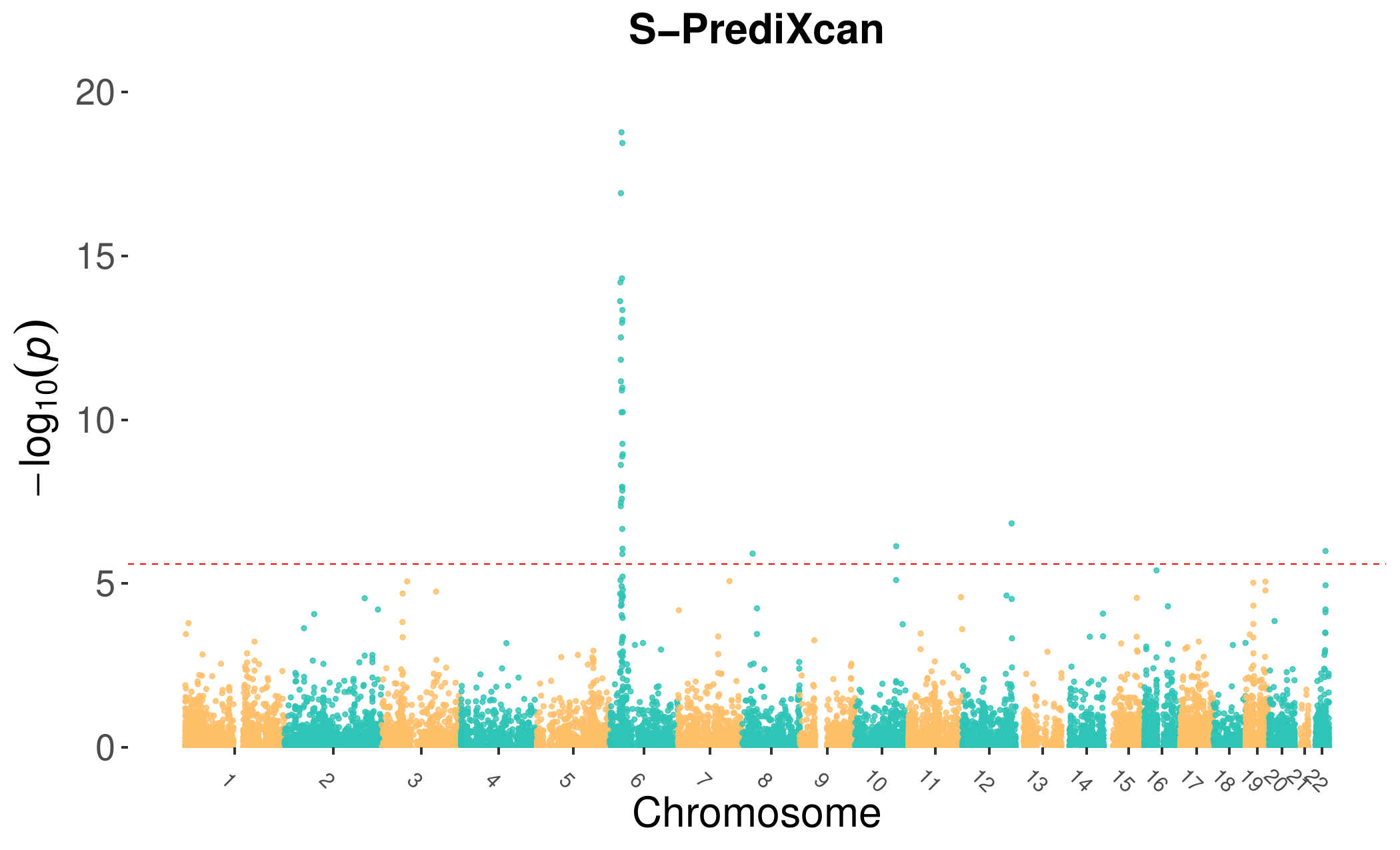}
		\caption{Manhattan plot for SCZ with significance threshold 2.5$e-6$.}
		\label{fig:manhattan_SCZ}
	\end{figure}

		We applied both QTWAS and S-PrediXcan to summary statistics from 
		a GWAS on schizophrenia \cite{pardinas2018common}, which included 
		11,260 cases and 24,542 
		controls. We use 13 brain tissues from GTEx data, with sample sizes 
		ranging from 114 to 205. The $p$-values from the 13 brain tissues are 
		combined by the Cauchy method for the final $p$-value. QTWAS 
		identified 76 genes, and S-PrediXcan identified 33 genes, with 18 genes 
		overlapping between 
		the two methods (Figure \ref{fig:manhattan_SCZ}). We performed a similar 
		enrichment analysis 
		as for LDL, identifying 17 out of 58 genes as enriched in the top five traits, 
		including autism and schizophrenia and complement C4, which is 
		genetically and neurobiologically related to schizophrenia (Table 
		\ref{tab:enrich2}). We also include the analysis for the other 7 binary traits 
		in Appendix Section 4.
	
	\begin{table}[!ht]
		\centering
		\resizebox{\textwidth}{!}{
		\begin{tabular}{lcc}
			\hline
			Trait& q-value (Bonferroni)& Number of Gene Hits\\ 
			\hline
			autism spectrum disorder, schizophrenia & 9.507e-10 &  12 \\ 
			complement C4 measurement & 8.927e-09 &  7 \\ 
			Takayasu arteritis & 4.392e-07 & 8\\ 
			ubiquitin carboxyl-terminal hydrolase 25 measurement & 6.755e-07 
			&  
			5 \\ 
			Epstein Barr virus nuclear antigen 1 IgG measurement   & 1.938e-06 
			&  
			4 \\ 
			
			\hline
		\end{tabular}
	}
		\caption{Enrichment analysis of the top 5 traits associated with genes 
			uniquely identified by QTWAS for schizophrenia. }
		\label{tab:enrich2}
	\end{table}

	\section{Discussion}\label{sec:discussion}

	We have proposed a novel quantile-based TWAS approach that utilizes a 
	conditional quantile process to model heterogeneous gene expressions 
	and combines it with GWAS summary statistics to infer nonlinear gene-trait 
	associations. Our framework is particularly useful in scenarios where gene 
	expression 
	patterns and underlying genetic regulation exhibit significant 
	heterogeneity, allowing for more flexible and precise modeling of gene-trait 
	associations. As demonstrated in both simulations and applications, our 
	quantile-based 
	TWAS is able to identify more gene-trait associations than traditional 
	methods and 
	provide deeper insights regarding how gene expression levels regulate 
	phenotypes, especially when their relationships vary across different 
	quantiles of gene expression. These findings underscore the potential of our 
	approach to enhance the discovery of genetic contributions to complex traits 
	and 
	diseases.
	

		Through validation analyses, we show that the novel genes identified by 
		QTWAS are likely to be functional and relevant to the trait under study.   
		There are, however, several sources of confounding that lead to false 
		positive associations in TWAS analyses. 
		First,  LD confounding and co-regulation 
		\citep{wainberg2019opportunities} can lead to false positive associations 
		and fine-mapping methods can further prioritize relevant genes at each 
		locus 
		\citep{ma2023bigknock,ma2021powerful,mancuso2019probabilistic}. 
		Second, QTWAS, as described here, focuses on estimating the 
		association between genetically predicted 
		gene expression and traits, but there are no guarantees of causal inference. 
		Colocalization methods can identify genetic variants that are 
		causal for two phenotypes, including gene expression and a trait of 
		interest \citep{zuber2022combining}. Examples include Bayesian 
		approaches and likelihood-based methods 			
		\citep{giambartolomei2014bayesian,plagnol2009statistical}. Similarly, 
		Mendelian Randomization (MR) methods have been proposed to test for 
		causal relationships between gene 
		expression and traits, by treating  eQTLs as the instrumental valuables 
		\citep{shi2020tissue}. A thorough review of the connections between 
		TWAS and MR can be found in \cite{zhu2021transcriptome}.
		Future research will be needed to adapt MR to quantile 
		association tests.  Furthermore, violations of assumptions, such as no 
		horizontal pleiotropy, can also lead to false positives for both QTWAS and 
		MR-based approaches. Increasing efforts have been made to adjust 
		TWAS for horizontal 
		pleiotropy 			
		\citep{deng2021model,zhao2024adjusting,gleason2021robust}. We 
		provide additional details on how horizontal pleiotropy could be included 
		in QTWAS in future work in Appendix Section 6. 
	
	
	In addition, several emerging topics are 
	worth exploring for future work. It is desirable to develop 
	approaches that combine data on multiple tissues to increase the total 
	sample size of eQTL studies for the estimation of the conditional distribution 
	of gene 
	expression. Such approaches have been developed before, e.g., 
	UTMOST \citep{hu2019statistical}, and have been shown to effectively 
	increase 
	imputation accuracy and power. Multi-tissue quantile modeling may allow 
	investigations of more comprehensive nonlinear associations across tissues. 
	Furthermore, 
	the current QTWAS framework can be better developed when individual 
	GWAS data are available, which would allow nonparametric approaches to 
	estimate 
	higher-resolution nonlinear gene-trait associations and multiplicative 
	errors instead of additive errors, depending on the original transformation 
	applied to the trait.  
	

		\section*{Software}QTWAS has been implemented in publicly 
		available software. We have posted a comprehensive demonstration on
		Github: https://github.com/tianyingw/QTWAS, which contains the 
		tutorial for downloading and using the pre-trained QTWAS models to 
		conduct analysis. 
	
	\section*{Acknowledgements} We would like to thank the Editor, 
	Associate Editor, and three anonymous referees for their valuable 
	comments and suggestions. The GTEx Project was supported by the 
	Common Fund of the Office of the Director of the National Institutes of 
	Health, and by NCI, NHGRI, NHLBI, NIDA, NIMH, and NINDS.
	This study was supported by the National Institutes of Health grants 
	AG087496 and AG072272.

	\bibliographystyle{chicago}
	\bibliography{references2}
	
\end{document}